\documentclass[12pt]{article}

\usepackage{amssymb,amsmath,bbm}
\usepackage{graphicx}
\usepackage{fullpage}
\usepackage{color}
\usepackage{hyperref}

\usepackage{amssymb,amsmath}

\renewcommand{\a}{\alpha}
\renewcommand{\b}{\beta}
\newcommand{\g}{\gamma}
\renewcommand{\d}{\delta}\newcommand{\D}{\Delta}

\newcommand{\z}{\zeta}

\renewcommand{\l}{\lambda}
\newcommand{\m}{\mu}

\renewcommand{\r}{\rho}
\renewcommand{\S}{\Sigma}

\renewcommand{\O}{\Omega}

\newcommand{\cN}{\mathcal{N}}
\newcommand{\cO}{\mathcal{O}}

\newcommand{\IC}{\mathbb{C}}

\newcommand{\IF}{\mathbb{F}}

\newcommand{\IP}{\mathbb{P}}

\newcommand{\IR}{\mathbb{R}}

\newcommand{\IT}{\mathbb{T}}

\newcommand{\IZ}{\mathbb{Z}}

\newcommand{\quotient}[1]{_{\hskip-2pt\lower1pt\hbox{$/$}\lower2pt\hbox{\hskip-1pt$#1$}}}
\newcommand{\ii}{\text{i}}

\def\place#1#2#3{\vbox to0pt{\kern-\parskip\kern-7pt
                             \kern-#2truein\hbox{\kern#1truein #3}
                             \vss}\nointerlineskip}

\newcommand{\sref}[1]{Section~\ref{#1}}

\newcommand{\fref}[1]{Figure~\ref{#1}}

\newcommand{\eref}[1]{Equation~\eqref{#1}}

\newcommand{\hodgenos}{(h^{1,1},\,h^{2,1})}
\newcommand{\symm}[1]{_{\hskip-3pt\lower3pt\hbox{$\left\{#1\right\}$}}}

\newcommand{\cicystop}{~\lower8pt\hbox{.}}

\def\place#1#2#3{\vbox to0pt{\kern-\parskip\kern-7pt
                             \kern-#2truein\hbox{\kern#1truein #3}
                             \vss}\nointerlineskip}

\newcommand{\beq}{\begin{equation}}
\newcommand{\eeq}{\end{equation}}
\newcommand{\bea}{\begin{eqnarray}}
\newcommand{\eea}{\end{eqnarray}}
\newcommand{\bean}{\begin{eqnarray*}}
\newcommand{\eean}{\end{eqnarray*}}

\newcommand{\comment}[1]{}

\newcommand{\+}{\phantom{-}}

\begin{document}
\renewcommand{\baselinestretch}{1.1}

\begin{center}
{\LARGE Hyperconifold Transitions, Mirror Symmetry,\\
and String Theory\\}
\vspace{.5in}
Rhys Davies\footnote{\it daviesr@maths.ox.ac.uk} \\
\vspace{.15in}
{\it
Mathematical Institute, \\
University of Oxford, \\
24-29 St Giles, Oxford \\
OX1 3LB, UK}
\end{center}
%\vspace{.5in}

\abstract{
Multiply-connected Calabi-Yau threefolds are of particular interest for both string
theorists and mathematicians.  Recently it was pointed out that one of the
generic degenerations of these spaces (occurring at codimension one in moduli
space) is an isolated singularity which is a finite cyclic quotient of the conifold;
these were called hyperconifolds.  It was also shown that if the order of the quotient
group is even, such singular varieties have projective crepant resolutions, which
are therefore smooth Calabi-Yau manifolds.  The resulting topological transitions
were called hyperconifold transitions, and change the fundamental group as well
as the Hodge numbers.  Here Batyrev's construction of Calabi-Yau hypersurfaces
in toric fourfolds is used to demonstrate that certain compact examples containing
the remaining hyperconifolds --- the $\IZ_3$ and $\IZ_5$ cases --- also have
Calabi-Yau resolutions.  The mirrors of the resulting transitions are studied and it is
found, surprisingly, that they are ordinary conifold transitions.  These are the first
examples of conifold transitions with mirrors which are more exotic extremal
transitions.  The new hyperconifold transitions are also used to construct a small
number of new Calabi-Yau manifolds, with small Hodge numbers and fundamental
group $\IZ_3$ or $\IZ_5$.  Finally, it is demonstrated that a hyperconifold is a physically
sensible background in Type IIB string theory.  In analogy to the conifold case,
non-perturbative dynamics smooth the physical moduli space, such that hyperconifold
transitions correspond to non-singular processes in the full theory.
}

\newpage

\tableofcontents

%%%
\section{Introduction and discussion}\label{sec:intro}

This paper is a follow-up to \cite{Davies:2009ub}, in which a class of
threefold singularities and associated topological transitions were studied.
These are isolated Calabi-Yau threefold singularities which are quotients of the
conifold by a finite cyclic group $\IZ_N$; such a singularity was named a
$\IZ_N$-hyperconifold.  They occur naturally in singular Calabi-Yau varieties
which are limits of families of smooth multiply-connected spaces, when the
generically-free group action on the covering space develops a fixed point.

It was shown in \cite{Davies:2009ub} that any projective variety with a
$\IZ_{2M}$-hyperconifold singularity has a projective crepant resolution,
establishing the existence of hyperconifold \emph{transitions} between smooth
compact Calabi-Yau threefolds.  The analysis was not sufficient to demonstrate
the existence of the remaining cases --- the $\IZ_3$- and $\IZ_5$-hyperconifold
transitions --- as the local resolution process did not guarantee that the resolved
manifold was projective (and hence K\"ahler).  Like the more familiar conifold
transitions, hyperconifold transitions change the Hodge numbers; for a
$\IZ_N$-hyperconifold transition, the change is
\begin{equation} \label{eq:hodgenos}
    \d\hodgenos_{\IZ_N} = (N-1, -1)~.
\end{equation}
A novel feature is that the fundamental group can also change.

The present work has several objectives.  We work mainly within the class
of Calabi-Yau hypersurfaces in toric fourfolds, first described systematically by
Batyrev \cite{Batyrev:1994hm} and then enumerated by Kreuzer and Skarke
\cite{Kreuzer:2000xy}.  The formalism is reviewed in \sref{sec:toricreview}, and
then used in \sref{sec:examples} to demonstrate that $\IZ_3$- and $\IZ_5$-hyperconifold
transitions do connect compact Calabi-Yau manifolds.  Perhaps more interestingly,
it can also be used to study the mirror processes to these transitions, which
turn out to be ordinary conifold transitions.  They therefore provide a counter-example
to an old conjecture of Morrison \cite{Morrison:1995km} that the mirror of a conifold
transition is another conifold transition.  The examples herein show that, while this
is a very tempting conjecture, it is not true in general.  They also motivate a modest
conjecture, that the mirror process to any $\IZ_N$-hyperconifold transition is a
conifold transition in which the intermediate variety has $N$ nodes.  It is probably
possible to use the local techniques of \cite{Chiang:1999tz,Gross:2000} to prove this
\cite{Gross:priv}.

The mirror conifold transitions have another interesting feature.  Batyrev and Kreuzer
showed that within the class of Calabi-Yau hypersurfaces in toric fourfolds, mirror
symmetry exchanges the fundamental group (which in these cases can only be
$\IZ_2, \IZ_3$ or $\IZ_5$) with the Brauer group, which is the torsion part of $H^3(X,\IZ)$
\cite{Batyrev:2005jc}.  Since the hyperconifold transitions studied here destroy the
fundamental group, their mirror conifold transitions should destroy the Brauer group.
This is not a new phenomenon (see for example \cite{GrossPavanelli}), but here mirror
symmetry gives a clear reason for it to occur.

Once we know that hyperconifold transitions exist, we can use them to try to construct
new Calabi-Yau manifolds.  This was mentioned in \cite{Davies:2009ub}, but no explicit
examples were given.  In \sref{sec:moreZ3} and \sref{sec:moreZ5}, we use the new
results of this paper to construct some previously unknown Calabi-Yau manifolds via
$\IZ_3$- and $\IZ_5$-hyperconifold transitions.

If two Calabi-Yau manifolds are mathematically connected by a topological transition,
we might ask whether the corresponding physical theories, obtained by compactifying
string theory on these spaces, are also smoothly connected.  It is shown in
\sref{sec:string} that the physical moduli space, at least in Type IIB string theory, is
perfectly smooth through a point corresponding to a hyperconifold transition.  The
story is very similar to that of a conifold transition, worked out in \cite{Greene:1995hu}.
The results of \cite{Davies:2009ub} and the present paper therefore have significant
implications for the connectedness of the moduli space of Calabi-Yau threefolds, and the
associated string vacua.  Soon after Reid suggested the idea that all threefolds with
$c_1 = 0$ may be connected by conifold transitions \cite{ReidFantasy}, this was shown to
be true for almost all known Calabi-Yau examples \cite{Green:1988bp,Green:1988wa}.
But conifold transitions cannot change the fundamental group, so this cannot be the whole
story.  Hyperconifold transitions then fill an important gap, since they still involve relatively
mild singularities, but can change the fundamental group as well as the Hodge numbers.
Whether conifold and hyperconifold transitions between them can connect all
Calabi-Yau threefolds is an interesting open question.

Before moving on, it may be helpful to illustrate the hyperconifold phenomenon by
considering a simple non-compact example.  Let the group $\IZ_2$ act on $\IC^4$
as follows:
\begin{equation*}
    (y_1, y_2, y_3, y_4) \to (-y_1, -y_2, -y_3, -y_4)~.
\end{equation*}
Then suppose we have a hypersurface $\widetilde X$ given by a polynomial
equation $f=0$.  If we want $\widetilde X$ to be invariant under $\IZ_2$,
and its quotient $X = \widetilde X/\IZ_2$ to be Calabi-Yau, then the
polynomial $f$ must be invariant.  As such, it can be written (perhaps after
a change of coordinates) as
\begin{equation}\label{eq:deformedconifold}
    f = \a_0 + y_1\, y_4 - y_2\, y_3 + \cO(y^3)~,
\end{equation}
since no invariant linear terms exist.  For $\a_0 \neq 0$, $\widetilde X$ is
smooth, and does not contain the origin, so the quotient $X$ is also smooth,
with fundamental group isomorphic to $\IZ_2$.  However, if we take the
limiting case $\a_0 = 0$, we see that $\widetilde X$ then contains the origin,
and that this point is a node, or conifold singularity.  The corresponding
singularity on $X$ is therefore a $\IZ_2$ quotient of the conifold.  Locally,
it looks like the vector bundle $\cO(-2,-2) \to \IP^1{\times}\,\IP^1$, with the
zero section projected to a point.  Blowing up the singular point gives a crepant
resolution of the singularity by restoring this zero section.  For more details,
including the toric data for this and the other hyperconifold singularities, see
\cite{Davies:2009ub}.

Our main interest here is in compact Calabi-Yau threefolds, and transitions
between them.  Most known multiply-connected Calabi-Yau threefolds are
obtained as free quotients of complete intersections in products of projective
spaces.  A few examples were discovered long ago
\cite{CHSW,Strominger:1985it,CICYS2}, and recently a more systematic
search has been performed, leading to a complete enumeration of the
manifolds which can be constructed this way
\cite{SHN,Braun:2010vc,Candelas:2010ve}.  A smaller number of
examples occur as hypersurfaces in toric fourfolds \cite{Batyrev:2005jc}, or as free
non-toric quotients of such hypersurfaces \cite{BCD}, which is a largely unexplored
class.\footnote{There are also certain exceptional cases, such as the quotients
of the Horrocks-Mumford quintic \cite{HorrocksMumford} and the Gross-Popescu
manifolds \cite{GrossPopescuI,GrossPopescuII}, but these are not discussed here.}
The cyclic fundamental groups which are known to occur are those of order
$N=2,3,4,5,6,8,10,12$.  In all cases, there is an action of $\IZ_N$ on the ambient
space which has fixed points, and these are missed by a generic member of the
family of embedded Calabi-Yau threefolds.  If such a threefold is deformed until it
does contain a fixed point, the quotient variety develops a hyperconifold
singularity.\footnote{It is possible for worse singularities to occur instead, because
the quadratic terms in the analogue of \eref{eq:deformedconifold} may always be
degenerate.  This does not seem to happen in products of projective spaces.}

%%%
\section{Toric geometry and the Batyrev construction}\label{sec:toricreview}

Here we will briefly review Batyrev's construction of Calabi-Yau hypersurfaces
in toric varieties \cite{Batyrev:1994hm}.  This will serve mainly to
establish notation, as several conventions have been used in the literature.
We will specialise to the case of Calabi-Yau threefolds in toric fourfolds.

Let $N$ be a lattice, $N \cong \IZ^4$, and $M$ its dual lattice.  It is
convenient to choose a basis for $N$, with corresponding dual basis for $M$, so
we can use coordinates.  Points of $N$ correspond to one-parameter subgroups of
the algebraic torus $\IT^4 = \big(\IC^*\big)^4$ via the map
\begin{equation*}
    N \ni (n_1, n_2, n_3, n_4) ~\mapsto~ \{(\l^{n_1}, \l^{n_2}, \l^{n_3}, \l^{n_4})
        ~\vert~ \l \in \IC^*\}~,
\end{equation*}
while points of $M$ correspond to monomials on $\IT^4$ considered as an
algebraic variety, via the map
\begin{equation}\label{eq:monmap}
    M \ni (m_1, m_2, m_3, m_4) ~\mapsto~ t_1^{m_1}t_2^{m_2}t_3^{m_3}t_4^{m_4}~.
\end{equation}
We will denote by $\chi^m$ the monomial associated to $m\in M$. 
The two lattices are naturally embedded in the vector spaces
$N_\IR = N\otimes_\IZ \IR$ and $M_\IR = M\otimes_\IZ\IR$, respectively.

Batyrev's construction begins with a polytope $\D$ in $M_\IR$, which
satisfies the following conditions:
\begin{itemize}
    \item
    The vertices of $\D$ are lattice points \emph{i.e.} they lie on
    $M\subset M_\IR$.
    \item
    The faces of $\D$ lie on hyperplanes of the form
    \begin{equation*}
        H_n = \{ m \in M_\IR ~\vert~ \langle m, n \rangle \geq -1 \}
    \end{equation*}
    where $n \in N$ is a primitive lattice vector.\footnote{A lattice vector is
    called primitive if it is the first lattice point on a ray.}  Note that this implies
    that $\D$ contains the origin as its unique interior point.
\end{itemize}
Such a $\D$ is called reflexive.  We also define the dual polytope
$\D^*\subset N_\IR$ by
\begin{equation*}
    \D^* = \{ n \in N_\IR ~\vert~ \langle m, n\rangle \geq -1 ~\forall~
        m \in \D \}~.
\end{equation*}
By taking cones over the faces of $\D^*$, we get the fan for a toric
variety which we will denote by $\IP_\D$ (the notation reflects the
fact that every variety constructed this way is projective).  It is a simple
fact that $\D^*$ is also reflexive.

We need one final definition.  Given a Laurent polynomial
$f = \sum_{m\in M} c_m \chi^m$, its Newton polytope is the convex
hull in $M_\IR$ of those points for which $c_m \neq 0$.  We will be
interested in those $f$ which have $\D$ as their Newton polytope.  The
vanishing of such an $f$ gives an affine sub-variety of $\IT^d$, and the
closure of this inside $\IP_\D$ is a Calabi-Yau variety.

Since both $\D$ and $\D^*$ are reflexive, we can reverse their roles in
the above construction.  The two families of Calabi-Yau hypersurfaces
are then mirror to each other.

\subsection{Homogeneous coordinates}

It is very convenient to use homogeneous coordinates for the ambient toric
space, as introduced by Cox in \cite{Cox:1993fz}.  Let $\S$ be a fan for
a toric variety $Z$.  Then we can construct $Z$ from $\S$ as follows.

Suppose $\S$ contains $d$ one-dimensional cones, which are rays, and let
$v_\r$ be the first lattice vector on the $\r$'th ray.  We associate with
it a complex coordinate $z_\r$.  Together, these are coordinates on $\IC^d$,
and will be our homogeneous coordinates for $Z$.  As in the construction of
ordinary projective space, our first step is to delete a certain subset of
$\IC^d$.  In short, we \emph{keep} the set where $z_{\r_1},\ldots,z_{\r_k}$
vanish simultaneously if and only if the vectors $v_{\r_1},\ldots,v_{\r_k}$
span a cone in $\S$.  We then impose a number of equivalence relations
on the resulting space, one for each linear relation satisfied by the vectors,
as follows
\begin{equation*}
    \sum_\r a_\r\, v_\r = 0 ~\Rightarrow~ (z_1,z_2,\ldots,z_d)
        \sim (\l^{a_1}z_1,\l^{a_2}z_2,\ldots,\l^{a_d}z_d) ~~\forall~~ \l\in\IC^*~.
\end{equation*}
There can be further, discrete identifications, which will be important for us,
but we will postpone their discussion for now.

In the cases of interest, $\S$ consists of cones over (some triangulation of)
the faces of a reflexive polytope $\D^*$.  Calabi-Yau hypersurfaces can
now be defined by the vanishing of homogeneous polynomials, which are
obtained from points of $M$ via a homogeneous version of
\eref{eq:monmap}:
\begin{equation}\label{eq:homogmonmap}
    M \ni m ~\mapsto~ \prod_{\r} z_\r^{\langle m, v_\r \rangle+1}~.
\end{equation}

%%%
\section{Transitions between toric hypersurfaces and their quotients}\label{sec:examples}

In this section we will turn to examples of hyperconifold transitions between
Calabi-Yau hypersurfaces in toric fourfolds.  The required analysis of reflexive
polytopes was greatly assisted by the software package PALP
\cite{Kreuzer:2002uu}.

\subsection
[The Z3 quotient of the bicubic]
{The $\IZ_3$ quotient of the bicubic} \label{sec:bicubic}

The family of `bicubic' manifolds $X^{2,83}$ are hypersurfaces in
$\IP^2\!\times\IP^2$, cut out by a single polynomial of bidegree
$(3,3)$.  Products of projective spaces are toric varieties, so we can use
Batyrev's formalism for Calabi-Yau hypersurfaces in toric fourfolds
\cite{Batyrev:1994hm}.

If we take homogeneous coordinates $(z_0,z_1,z_2)$ on the first
$\IP^2$ and $(z_3,z_4,z_5)$ on the second $\IP^2$, then we can take the
corresponding vectors in $N\cong\IZ^4$ to be
\begin{equation*}
    \IP^2{\times}\,\IP^2:\hskip20pt
    \begin{array}{rrrrrr}
    z_0 & z_1 & z_2 & z_3 & z_4 & z_5 \\[1ex]\hline
    1 & 0 & -1 & 0 & 0 & 0 \\
    0 & \+1 & -1 & 0 & 0 & 0 \\
    0 & 0 & 0 & \+1 & 0 & -1 \\
    0 & 0 & 0 & 0 & \+1 & -1 \end{array}~~\raisebox{-30pt}{.}
\end{equation*}
It is easy to see that the linear relations between these vectors induce the
two expected rescalings of the coordinates.  The convex hull of these six
points is a reflexive polytope~$\D^*$.

Using \eref{eq:homogmonmap}, we can write down the monomial corresponding
to a point of the dual lattice $M$ in the present case:
\begin{equation}\label{eq:bicubicmonos}
    (m_1, m_2, m_3, m_4) ~\mapsto~ z_0^{1+m_1}z_1^{1+m_2}z_2^{1-m_1-m_2}
    					   z_3^{1+m_3}z_4^{1+m_4}z_5^{1-m_3-m_4}
					   \equiv \chi^m ~.
\end{equation}
It is easy enough to check that the polytope $\D$, dual to $\D^*$ above,
corresponds exactly to bicubic monomials under this map.

We now define an action of $\IZ_3$, generated by
\begin{equation}\label{eq:firstZ3}
    g_3 ~:~ z_i \to \z^i\,z_i ~,
\end{equation}
where $\z = \exp(2\pi\ii/3)$.  The resulting orbifold
$(\IP^2\times\IP^2)/\IZ_3$ is also toric, and we obtain its fan simply by
sub-dividing the lattice $N$.  It is instructive to carry this out explicitly.
Polynomials on the quotient are exactly those polynomials on the covering
space which are invariant under the $\IZ_3$ action.
Under this action, we see from \eqref{eq:bicubicmonos} that
\begin{equation*}
    \chi^m ~\to~ \z^{m_1-m_2+m_3-m_4}\,\chi^m~,
\end{equation*}
so the sub-lattice $M' \subset M$ corresponding to $\IZ_3$-invariants
is determined by the condition $m_1-m_2+m_3-m_4 \equiv 0$ mod
3.  The polytope $\D$ is also reflexive with respect to $M'$, and so
determines a family of Calabi-Yau hypersurfaces in the quotient.

A short algebraic exercise determines a basis for the corresponding dual
lattice $N' \subset N_\IR$, which is a refinement of the lattice $N$:
\begin{equation*}
    N' = \left\langle
        (1,0,0,0), (0,1,0,0), (0,0,1,0), \left(\!-\frac13,\frac13,-\frac13,\frac13\right)
        \right\rangle~.
\end{equation*}
We can re-express the generators of our fan in terms of this basis:
\begin{equation*}
     (\IP^2{\times}\,\IP^2)/\IZ_3:\hskip20pt
    \begin{array}{rrrrrr}
    z_0 & z_1 & z_2 & z_3 & z_4 & z_5 \\[1ex]\hline
    1 & 0 & -1 & 0 & 1 & -1 \\
    0 & \+1 & -1 & 0 & -1 & 1 \\
    0 & 0 & 0 & \+1 & 1 & -2 \\
    0 & 0 & 0 & 0 & 3 & -3 \end{array}~~\raisebox{-30pt}{.}
\end{equation*}

Generic hypersurfaces determined by $\D$ miss the orbifold points, and
therefore give a family of smooth, multiply connected Calabi-Yau
threefolds, with Hodge numbers $\hodgenos = (2,29)$.  This is well known;
see for example \cite{Triadophilia}.  We are now interested in specialising
to the case where the Calabi-Yau hypersurface intersects one of these
singularities, and therefore has a $\IZ_3$-hyperconifold singularity.

Let us focus on the fixed point $z_1 = z_2 = z_4 = z_5 = 0$.  There is a single
monomial which is non-zero at this point: $z_0^3z_3^3$.  This 
corresponds to the point $(2,-1,2,-2) \in M'$, which is in fact a vertex of
$\D$.  So we obtain a hyperconifold singularity precisely when the
coefficient of this monomial vanishes, and in this case the polynomial
$f$ has a reduced Newton polytope $\widehat\D$, obtained as the convex
hull of the lattice points in $\D$, minus the vertex above.

So it is now clear how the resolution process proceeds:  the new dual polytope,
$\widehat\D^*$, is larger than $\D^*$, and the extra vertices correspond to
exceptional divisors resolving the orbifold singularity in the ambient space.  The
results of \cite{Batyrev:1994hm} imply that this resolves the $\IZ_3$-hyperconifold
singularity as well, but we will check this explicitly below.  We see that in this
case, and indeed all those in the present paper, the hyperconifold transition is a
link in the web of toric hypersurfaces described by Kreuzer and Skarke
\cite{Kreuzer:2000xy}.

\begin{figure}[!ht]
\begin{center}
    \includegraphics[width=.15\textwidth]{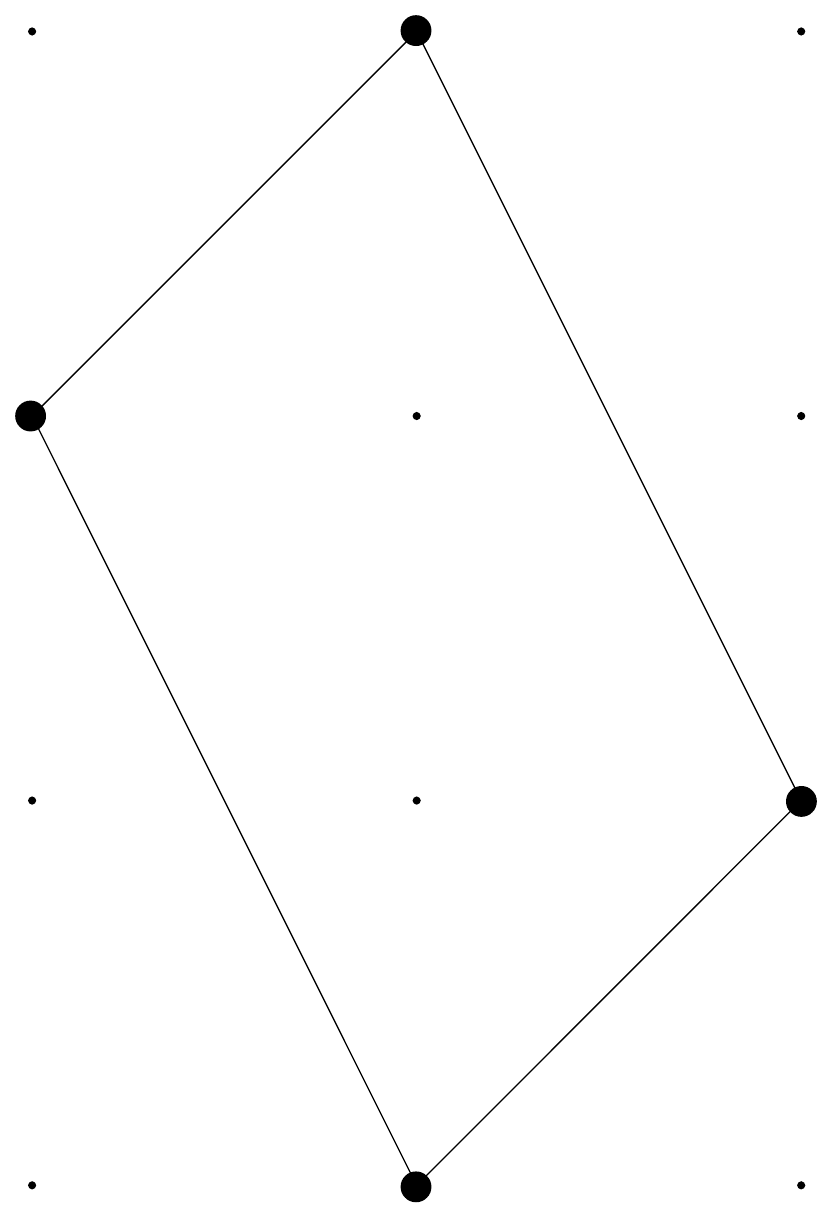}
    \hskip45pt
    \includegraphics[width=.15\textwidth]{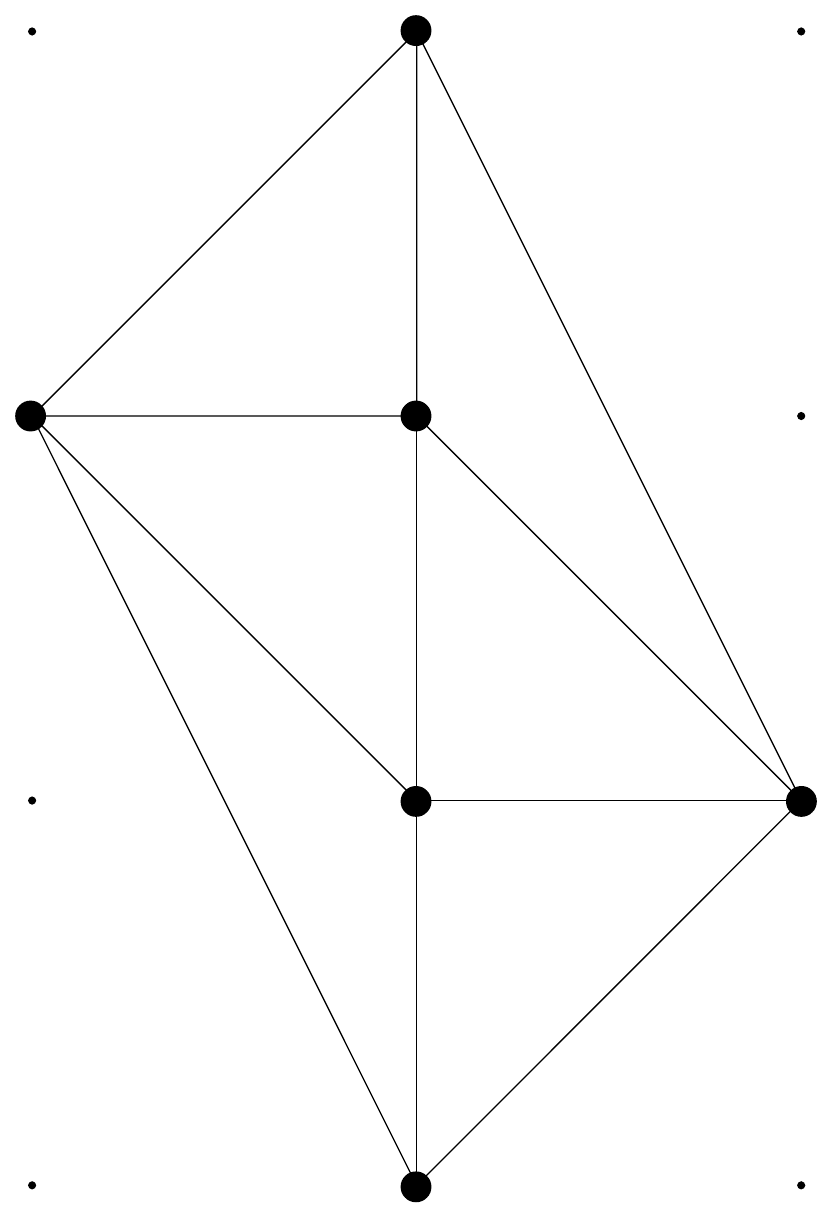}
    \hskip45pt
    \includegraphics[width=.15\textwidth]{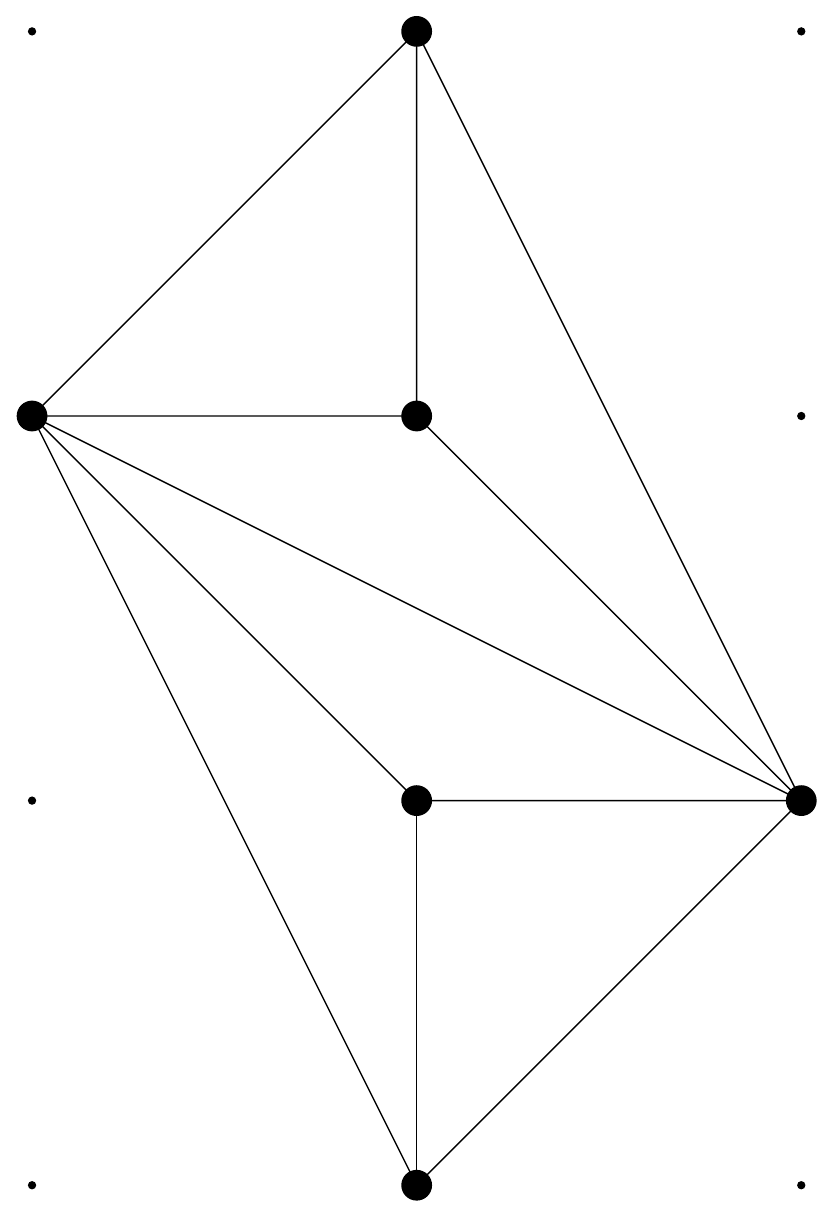}
    \\
\parbox{.75\textwidth}
{\caption{\label{fig:Z3hyperconifold}
\small The toric diagram for the $\IZ_3$-hyperconifold, and its two
crepant resolutions.  The first resolution is the one which occurs in
the example of this section.}}
\end{center}
\end{figure}

In \cite{Davies:2009ub},
the local toric structure of the $\IZ_3$-hyperconifold singularity was described,
and the corresponding toric diagram is reproduced in \fref{fig:Z3hyperconifold},
along with those for the two distinct local crepant resolutions.  The resolution we
have just implicitly constructed must correspond to one of these.  It will turn out to
be the first, but to see this we will have to go into more detail.

It turns out that $\widehat\D^*$, which corresponds to a space in
which the orbifold singularity is resolved, is obtained from $\D^*$ by
adding two more one-dimensional cones, which are contained in the
top-dimensional cone of $\D^*$ corresponding to the orbifold
point.\footnote{It should be noted that adding just one of the two new
cones gives a polytope which is not reflexive; there is no `halfway house'
between $\D^*$ and $\widehat\D^*$.}
We will call the corresponding new homogeneous coordinates
$z_6, z_7$; our list of coordinates, and corresponding lattice points, is
now
\begin{equation*}
    \begin{array}{rrrrrrrr}
    z_0 & z_1 & z_2 & z_3 & z_4 & z_5 & z_6 & z_7 \\[1ex]\hline
    1 & 0 & -1 & 0 & 1 & -1 & -1 & 0 \\
    0 & \+1 & -1 & 0 & -1 & 1 & 0 & 0 \\
    0 & 0 & 0 & \+1 & 1 & -2 & -1 & 0 \\
    0 & 0 & 0 & 0 & 3 & -3 & -1 & 1 \end{array}~~\raisebox{-30pt}{.}
\end{equation*}
The new relations are
\begin{eqnarray*}
    3v_6 - v_1 - 2v_2 - v_4 - 2v_5 = 0 \\[1ex]
    3v_7 - 2v_1 - v_2 - 2v_4 - v_5 = 0
\end{eqnarray*}

\begin{figure}[!ht]
\begin{center}
    \includegraphics[width=.25\textwidth]{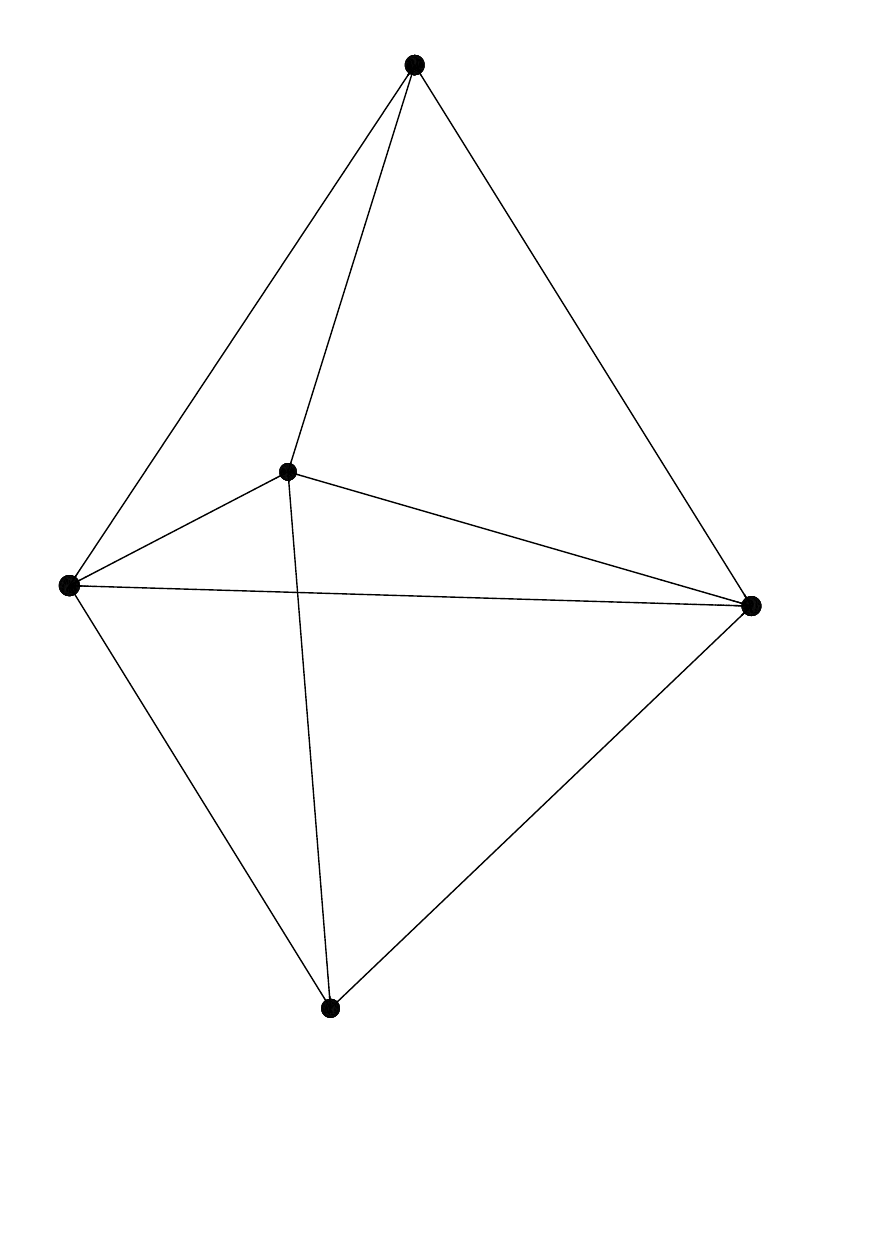}\\
    \place{3.0}{.4}{$z_0$}
    \place{3.1}{2.5}{$z_7$}
    \place{3.05}{1.63}{$z_1$}
    \place{2.38}{1.4}{$z_4$}
    \place{3.86}{1.37}{$z_5$}
    \vskip-25pt
    \parbox{.75\textwidth}
    {\caption{\label{fig:bipyramid}
    \small One of the non-simplicial faces of $\widehat\D^*$.  Adding the
    two-simplex $\langle z_1 z_4 z_5\rangle$ gives a maximal triangulation.  Vertices
    are labelled by the corresponding homogeneous coordinates.}}
\end{center}
\end{figure}

PALP gives us the various faces of $\widehat\D^*$.  There are four
non-simplicial facets, which must be triangulated in order to resolve the
corresponding toric fourfold.  We will focus on one such facet; the other
three can be treated identically.  Its vertices correspond to the
homogeneous coordinates $z_0, z_1, z_4, z_5, z_7$.  The
two-dimensional faces of this polyhedron are then
\begin{equation*}
    \begin{aligned}
        \langle z_0 z_1 z_4\rangle~&,&~\langle z_0 z_1 z_5\rangle~&,&~\langle z_0 z_4 z_5\rangle~,\\
        \langle z_1 z_4 z_7\rangle~&,&~\langle z_1 z_5 z_7\rangle~&,&~\langle z_4 z_5 z_7\rangle~.
    \end{aligned}
\end{equation*}
We see that $z_0$ and $z_7$ appear thrice each, while the other
coordinates each appear four times; this implies that the polyhedron
looks like \fref{fig:bipyramid}.  It has an obvious maximal triangulation,
given by adding a new two-dimensional face
$\langle z_1 z_4 z_5\rangle$, which divides it into two minimal
tetrahedra.  In fact, we have no choice but to take this triangulation --- we
are resolving an orbifold point of $(\IP^2\times\!\IP^2)/\IZ_3$, in which
$z_1, z_4, z_5$ are certainly allowed to vanish simultaneously, so this
two-face was already there.

Batyrev tells us that the procedure above resolves the
$\IZ_3$-hyperconifold singularity, and we would like to know to which
local resolution this corresponds, where the two possibilities are shown
in \fref{fig:Z3hyperconifold}.  To answer this question we will examine the
exceptional set of the resolution.  Inspection of \fref{fig:Z3hyperconifold},
and the `star construction' of toric geometry, as described in
\cite{Fulton}, tell us that in the first case, the exceptional set consists
of two copies of the Hirzebruch surface $\IF_1$, intersecting along a
$\IP^1$, while in the second it consists of two disjoint surfaces, each
isomorphic to $\IP^2$.

The two components of the exceptional set in the case at hand are given
by $z_6 = 0$ and $z_7 = 0$, respectively.  Let us examine the component
$z_6 = 0$ first.  After the triangulation described above, $z_6 = 0$
implies that $z_0 \neq 0$ and $z_3 \neq 0$.  We can therefore set
$z_0 = z_3 = 1$, using the usual rescaling relations of the two $\IP^2$'s.
This leaves us with homogeneous coordinates $z_1, z_2, z_4, z_5, z_7$
for some toric threefold, and remaining identifications which are equivalent
to the following:
\begin{equation*}
    (z_1, z_2, z_4, z_5, z_7) \sim
    (\m\,z_1, \l\,z_2, \m\,z_4, \l\,z_5,\m^{-2}\l\,z_7)~~,~\l,\m~\in~\IC^*~.
\end{equation*}
The interpretation of this is that $z_1, z_4$ are homogeneous coordinates
for a base $\IP^1$, while $z_2, z_5, z_7$ are homogeneous coordinates on
the fibres of the projective bundle
$\IP\big(\cO_{\IP^1}{\oplus}\cO_{\IP^1}{\oplus}\cO_{\IP^1}(-2)\big)$
(indeed, a careful inspection of the fan reveals that before taking the
quotient, we must delete the sets $\{z_1 = z_4 = 0\}$ and
$\{z_2 = z_5 = z_6 = 0\}$).  The exceptional divisor in our Calabi-Yau
hypersurface is then given by restricting the equation $f=0$ to this threefold.

If we take all the monomials coming from $\widehat\D$ and set
$z_0 = z_3 = 1$ and $z_6 = 0$, we are left with
\begin{equation*}
    \begin{array}{c}
    z_1 z_2~,~z_2 z_4~,~z_1 z_5~,~z_4 z_5~,~z_1^3 z_7~,~\\[1ex]
    z_1^2 z_4 z_7~,~z_1 z_4^2 z_7~,~z_4^3 z_7~.\end{array}
\end{equation*}
Consider an arbitrary linear combination of these.  If we set $z_1, z_4$
to any values, we are left with something linear in the homogeneous
coordinates of the $\IP^2$ fibre.  So the exceptional divisor is a $\IP^1$
bundle over the base $\IP^1$ parametrised by $z_1, z_4$, i.e. it is a
Hirzebruch surface.  An identical analysis holds for the component
$z_7 = 0$, and it is easily checked that the two components overlap on a
$\IP^1$, so the resolution realised is the first of those in
\fref{fig:Z3hyperconifold}.

Finally, we ask about the topological data of the resolved Calabi-Yau.  This
can be calculated directly from the polytope $\widehat\D$, and the
Hodge numbers turn out to be $\hodgenos = (4,28)$.  So the change
realised by the hyperconifold transition is $\d\hodgenos = (2,-1)$, in
accord with the argument of \cite{Davies:2009ub} which implied \eref{eq:hodgenos}
--- we imposed a single condition on the complex structure, and the resolution
introduced two new divisor classes.  Furthermore, the new family of manifolds
$X^{4,28}$ are simply-connected, because the fundamental group was destroyed
by allowing a fixed point of the $\IZ_3$ action to develop; this also follows simply
from Theorem 1.6 of \cite{Batyrev:2005jc}.

\subsubsection{The mirror transition}

Batyrev's construction allows us to easily identify the mirror of a Calabi-Yau
hypersurface in a toric variety:  we simply exchange the roles of the
polytopes $\D$ and $\D^*$.  This will allow us to identify the process
which is mirror to the above transition; on general grounds, it will be a projection
from $X^{29,2} \subset \IP_{\D^*}$ to a singular member of
$X^{28,4} \subset \IP_{\widehat\D^*}$, followed by a smoothing.  Surprisingly, we will
see that this turns out to be an ordinary conifold transition, as discussed in
\sref{sec:intro}.

We obtain $\widehat\D$ from $\D$ by removing a single vertex,
corresponding to blowing down a divisor.\footnote{The divisor, which is a
threefold, may be blown down to a curve or a surface, depending on the
chosen triangulation of $\widehat\D$.  We are interested in maximal
triangulations, in which case the divisor is blown down to a surface, as
we will see.}  Four
other points, which were interior to higher-dimensional faces of $\D$,
become vertices of $\widehat\D$.  We will use $w$ instead of $z$ for the
homogeneous coordinates in this section, to avoid confusion, and order them
so that these four are $w_1, w_2, w_3, w_4$; the corresponding
points are the vertices of a two-face.  In any maximal triangulation of
$\widehat\D$, this must be divided into two triangles, by adding to the fan
either $\langle w_1 w_2\rangle$ or $\langle w_3 w_4\rangle$.  We will
consider the first option.  Then adding the new point to pass to $\D$
corresponds to blowing up along the toric surface $S$ given by $w_1 = w_2 = 0$.
We see this by noting that, if we call the new coordinate $w_0$, the associated
vectors satisfy $u_0 - u_1 - u_2 = 0$.
\begin{figure}[ht]
\begin{center}
\includegraphics[width=.3\textwidth]{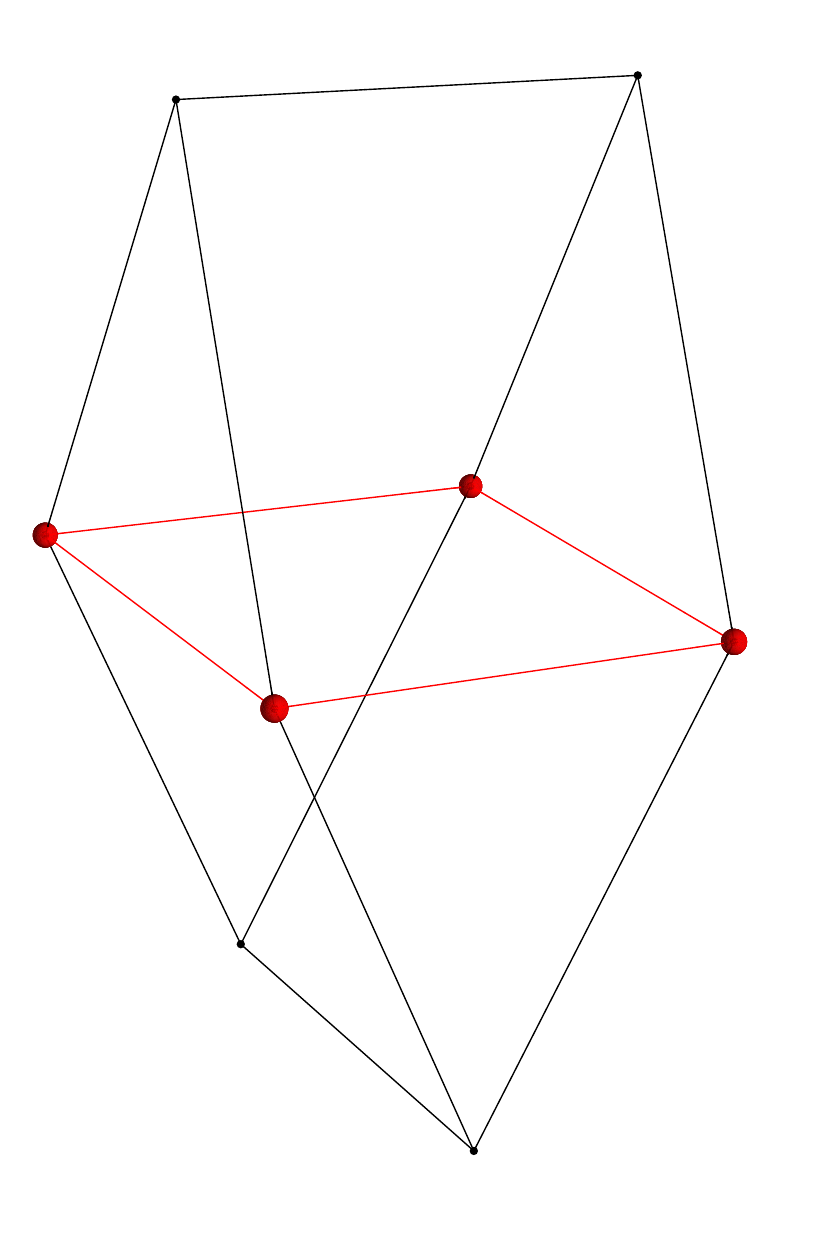}\\[2ex]
\place{3.4}{.4}{$w_8$}
\place{2.6}{1.0}{$w_7$}
\place{2.75}{1.45}{$w_1$}
\place{2.13}{2.0}{$w_4$}
\place{4.08}{1.65}{$w_3$}
\place{3.17}{2.17}{$w_2$}
\place{2.47}{3.09}{$w_5$}
\place{3.75}{3.17}{$w_6$}
\vskip-20pt
\parbox{.75\textwidth}
{\caption{\label{fig:biprism}
\small Part of the boundary of $\widehat\D$, with the vertices labelled
by the corresponding coordinates.  The red two-face can be
triangulated by adding the one-face $\langle w_1 w_2\rangle$.  This one-face
is then bisected by the new ray which is added to pass to $\D$.}}
\end{center}
\end{figure}

Now consider the singular Calabi-Yau varieties, which are mirror to the
hyperconifold from the last section.  These spaces are singular members of
$X^{28,4}$, given by setting to zero the coefficients of the two monomials coming
from the points $v_6, v_7$ of $\widehat\D^*$, which do not belong to $\D^*$.  It
is a quick check that these two monomials are the only ones which are not
identically zero on the surface $w_1 = w_2 = 0$, so the singular varieties contain
the surface $S$ from above, and are resolved when it is blown up.

\begin{figure}[ht]
\begin{center}
    \includegraphics[width=.5\textwidth]{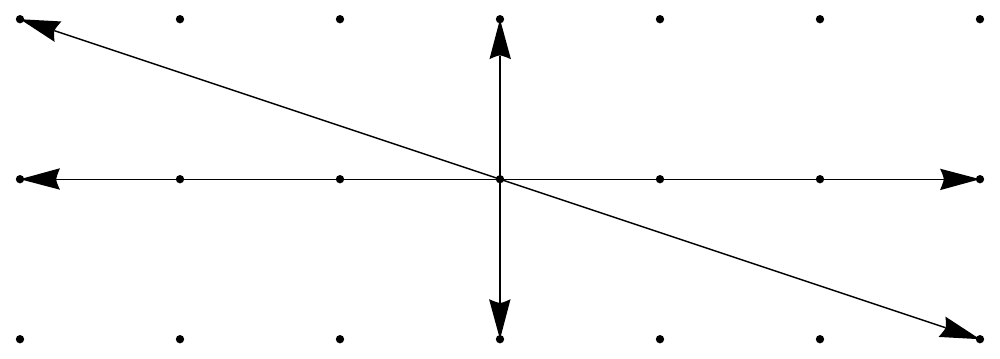}\\
    \place{4.85}{0.2}{$w_8$}
    \place{3.13}{0.13}{$w_7$}
    \place{1.45}{0.75}{$w_4$}
    \place{4.85}{0.75}{$w_3$}
    \place{1.45}{1.3}{$w_5$}
    \place{3.13}{1.4}{$w_6$}
    \vskip-10pt
    \parbox{.75\textwidth}
    {\caption{\label{fig:Z3Sfan}
    \small The fan for the toric surface $S$, along which we blow up to realise the
    conifold transition from $X^{28,4}$ to $X^{29,2}$.}}
\end{center}
\end{figure}

In the ambient toric space, the exceptional divisor is a $\IP^1$ bundle over
$S$, since $S$ is codimension two.  We want to calculate the exceptional set
in the resolved Calabi-Yau manifolds $X^{29,2}$.  After the introduction of
$w_0$, the coordinates $w_1$ and $w_2$ become homogeneous coordinates
on the $\IP^1$ fibres.  Inspecting \fref{fig:biprism}, we see that on the
exceptional divisor, given by $w_0 = 0$, we must have
$w_i \neq 0 ~\forall~i>8$, so we can use the toric scaling relations to set
$w_i = 1 ~\forall~i>8$.  Then, setting $w_0 = 0$, the most general
polynomial defining our Calabi-Yau hypersurface is
\begin{equation*}
    w_1(\a_1 w_5^3 w_4 + \a_2 w_8^3 w_3) +
        w_2(\a_3 w_7^3 w_4 + \a_4 w_8^3 w_3) = 0~.
\end{equation*}
We can now see that we actually have an ordinary conifold transition!  There
is a unique solution for the ratio $[w_1 : w_2]$ unless the two quantities
in brackets vanish simultaneously, in which case an entire copy of $\IP^1$
projects to the corresponding point of $S$.  Starting from the fan for $S$,
obtained from the star construction and shown in \fref{fig:Z3Sfan}, it is easy
enough to check that this occurs at three points, so the exceptional set is three
disjoint copies of $\IP^1$.  This implies that the change in the Euler number is
$\D\chi = 3*2 = 6$, which is consistent with the change in the Hodge numbers.

The above is a standard story for conifold transitions.  The singular varieties
contain the surface $S$ as a non-Cartier divisor, which passes through the
three nodes.  Blowing up along $S$ provides a small resolution of all the nodes,
and the resulting smooth variety is guaranteed to be projective.
So the mirror of the $\IZ_3$-hyperconifold transition is an ordinary conifold
transition, where the intermediate singular variety has three nodes.
Note that according to \cite{Batyrev:2005jc}, $X^{28,4}$ has no torsion in its
cohomology, whereas $X^{29,2}$ has Brauer group $\IZ_3$, so these conifold
transitions change the Brauer group.

\subsubsection{Multiple hyperconifolds} \label{sec:chain}

We started with Calabi-Yau hypersurfaces $X^{2,29}$ in $\IP^2{\times}\,\IP^2/\IZ_3$,
and saw that imposing a single condition on the complex structure causes them
to intersect one of the orbifold points.  Clearly, there is no reason why we cannot
do this for multiple points at once.  The resolution process is essentially local, so
we get transitions where the intermediate variety has multiple hyperconifold
singularities.

Alternatively, we can think of doing this in distinct steps:  after performing a single
hyperconifold transition to $X^{4,28}$, the ambient space still has a number of orbifold
points, and we can ask for the hypersurface to intersect one of these.  This process can
continue while the ambient space still has unresolved orbifold points.  There are nine
fixed points of the original $\IZ_3$ action on $\IP^2{\times}\,\IP^2$, so we get a chain of
nine hyperconifold transitions\footnote{One can check that at no point do any `extra'
singularities arise from restricting the complex structure.}
\begin{equation*}
    X^{2,29} \rightsquigarrow X^{4,28} \rightsquigarrow X^{6,27} \rightsquigarrow
        X^{8,26} \rightsquigarrow \ldots \rightsquigarrow X^{20,20}~.
\end{equation*}
Only at the first step is there any change in the torsion part of the (co)homology.

\subsubsection
[New manifolds from the Z3 x Z3' quotient]
{New manifolds from the $\IZ_3{\times}\IZ_3'$ quotient}\label{sec:moreZ3}

All the manifolds discussed above are hypersurfaces in toric fourfolds, and therefore
already appear in the Kreuzer-Skarke list; we merely showed that they are connected
by hyperconifold transitions.  Here we turn to an example of how new Calabi-Yau
manifolds can be constructed by considering hyperconifold transitions from known
ones.

A smooth sub-family of bicubics actually admit a free action by $\IZ_3{\times}\IZ_3'$,
giving a smooth quotient family with Hodge numbers $\hodgenos = (2,11)$.  The
first $\IZ_3$ still acts as in \eref{eq:firstZ3}, but the second does not act torically, instead
permuting the homogeneous coordinates of each $\IP^2$,
\begin{equation*}
    g_3' ~:~ z_0 \to z_1 \to z_2 \to z_0 ~,~~ z_3 \to z_4 \to z_5 \to z_3~.
\end{equation*}
The quotient manifolds $X^{2,11} = X^{2,83}/\IZ_3{\times}\IZ_3'$ are therefore not toric
hypersurfaces, and we will see that we can generate genuinely new manifolds by
hyperconifold transitions from them.

The action of $g_3'$ on the ambient space permutes the nine fixed points of
$g_3$, which therefore fall into three orbits of three.  So now, when we ask for a
fixed point of $g_3$ to develop on the covering space $X^{2,83}$, three will in fact
develop, and these are identified by the action of $g_3'$.  Taking the quotient by just
$\IZ_3$ and simultaneously resolving the three singularities will realise the transition
from $X^{2,29}$ to $X^{8,26}$, i.e. the first three links in the chain of last section, in one
step.  We can restrict the K\"ahler form such that the exceptional divisors over each point
have the same volume,  and in this way $X^{8,26}$ inherits a free action of $\IZ_3'$, by
which we can quotient.  We have therefore in fact described a hyperconifold transition from
$X^{2,11} = X^{2,83}/\IZ_3{\times}\IZ_3'$ to a new manifold $X^{4,10} = X^{8,26}/\IZ_3'$.
As before, we can now perform the same process for the remaining $\IZ_3'$-orbits of fixed
points, of which there are two.

In summary, we obtain a short chain of hyperconifold transitions,
\begin{equation*}
    X^{2,11} \rightsquigarrow X^{4,10} \rightsquigarrow X^{6,9}
        \rightsquigarrow X^{8,8}~,
\end{equation*}
where the last three spaces all have fundamental group $\IZ_3$, being free quotients
by $g_3'$ of $X^{8,26}$, $X^{14,23}$, $X^{20,20}$ respectively.  The families $X^{4,10}$
and $X^{8,8}$ are certainly new manifolds, since no existing manifolds have the same
Hodge numbers and fundamental group.  $X^{6,9}$, on the other hand, could well be
the same as Yau's famous three-generation manifold \cite{Yau1,Greene:1986bm}.  This
suspicion is strengthened by the fact that their covering spaces have the same Hodge
numbers.

\subsection
[The Z5 quotient of the quintic]
{The $\IZ_5$ quotient of the quintic}\label{sec:quintic}

A smooth quintic hypersurface in $\IP^4$ is a Calabi-Yau manifold, with Hodge
numbers $\hodgenos=(1,101)$.  If we take homogeneous coordinates
$z_i,~i=0,\ldots,4$, an action of $\IZ_5$ on $\IP^4$ can be defined by
\begin{equation*}
    g_5 ~:~ z_i \to \z^i z_i ~,
\end{equation*}
where $\z = \exp(2\pi\ii/5)$.  It is well known that generic quintic polynomials
invariant under this action determine smooth Calabi-Yau manifolds without fixed
points.  The resulting family of smooth quotients are $X^{1,21}$.  We can perform
an analysis almost identical to that in \sref{sec:bicubic} to show that there is a
hyperconifold transition to a simply-connected family $X^{5,20}$; here we only
sketch the details.

Imposing a single condition on the complex structure of $X^{1,21}$, we can arrange
for the covering space to contain one of the fixed points of the $\IZ_5$ action, say
$(1,0,0,0,0)$, and this gives rise to a $\IZ_5$-hyperconifold in $X^{1,21}$.  As in the
example of \sref{sec:bicubic}, the singular point is also a fixed point of the torus action
on the ambient space, and when we resolve it we get another toric fourfold.  The
resolution introduces four new coordinates in this case, which we label
$z_5, z_6, z_7, z_8$, such that the homogeneous coordinates and corresponding
vectors are:
\begin{equation*}
    \begin{array}{rrrrrrrrr}
    z_0 & z_1 & z_2 & z_3 & z_4 & z_5 & z_6 & z_7 & z_8 \\[1ex]\hline
    1 & 0 & 0 & -4 & 3 & -2 & 0 & -1 & 1 \\
    0 & \+1 & 0 & -3 & 2 & -1 & 0 & 0 & 1 \\
    0 & 0 & \+1 & -2 & 1 & -1 & 0 & 0 & 1 \\
    0 & 0 & 0 & 5 & -5 & 2 & -1 & 1 & -2 \end{array}~~\raisebox{-30pt}{.}
\end{equation*}
The proper transforms of the singular Calabi-Yau varieties are smooth Calabi-Yau
manifolds $X^{5,20}$ in this new ambient space.  The Hodge numbers follow from the
general formula \eqref{eq:hodgenos}, and are again confirmed by PALP.

\begin{figure}[ht]
\begin{center}
    \includegraphics[width=.35\textwidth]{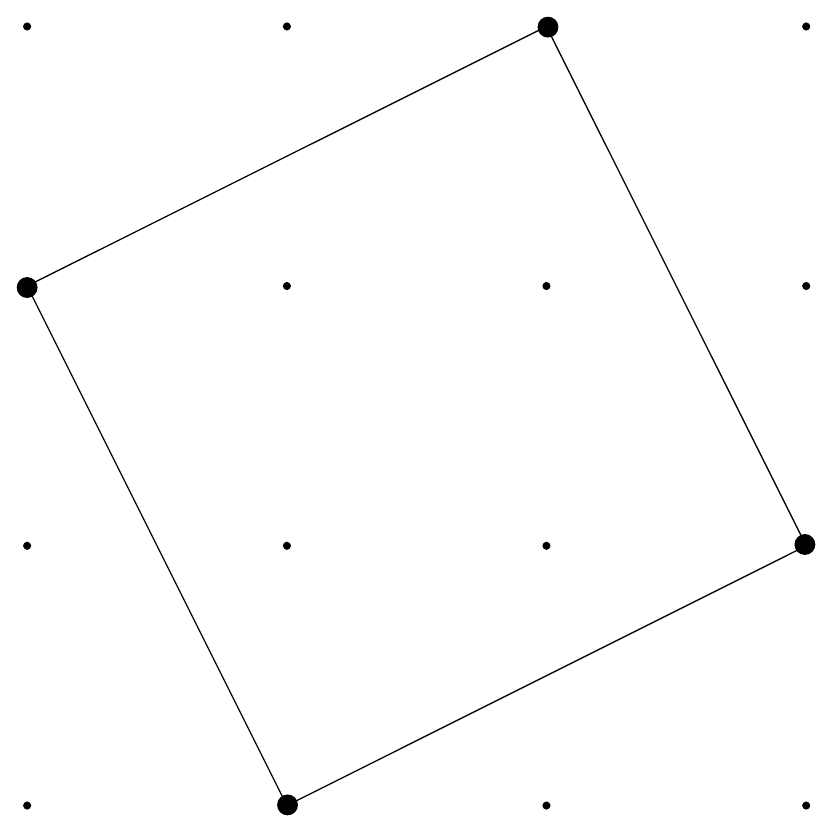}
    \parbox{.75\textwidth}{
    \caption{\label{fig:Z5hyperconifold}
    \small The toric diagram for the $\IZ_5$-hyperconifold.}
    }
\end{center}
\end{figure}

The toric diagram for the $\IZ_5$-hyperconifold singularity is shown in
\fref{fig:Z5hyperconifold}, and it is easy to see that there are several possible crepant
resolutions.  The topology of the one realised by the resolution constructed
here is more difficult to find than in the analogous problem of \sref{sec:bicubic}, and
has not been investigated.

The mirror to the transition above can be found by the same method as that in
\sref{sec:bicubic}, and the story is very similar.  The mirror varieties to the singular
members of $X^{1,21}$ are singular members of $X^{20,5}$, all of which contain a
toric surface $S'$, and generically have five nodes which lie on this surface.  Blowing
up along $S'$ resolves the nodes, leading to smooth manifolds $X^{21,1}$, and
this resolution is mirror to the deformation of the hyperconifold.  Similarly, the nodes
can be smoothed by passing to a general member of $X^{20,5}$, which is mirror to the
resolution of the hyperconifold.  So the mirror to this $\IZ_5$-hyperconifold transition is
a conifold transition in which the intermediate variety has five nodes, consistent with
our conjecture of \sref{sec:intro}.  Note that $X^{20,5}$ has no torsion in its cohomology,
but $X^{21,1}$ has Brauer group $\IZ_5$.

\subsubsection{More transitions, and another new manifold}\label{sec:moreZ5}

Again there are further hyperconifold transitions possible; this time there are five
orbifold points in the original ambient space $\IP^4/\IZ_5$, and we get the following chain
of transitions:
\begin{equation*}
    X^{1,21} \rightsquigarrow X^{5,20} \rightsquigarrow X^{9,19} \rightsquigarrow
        X^{13,18} \rightsquigarrow X^{17,17} \rightsquigarrow X^{21,16}~.
\end{equation*}
The first manifold has fundamental group $\IZ_5$, while the other five all have
torsion-free (co)homology.

In analogy with the bicubic case, we can now consider the action of a second group
$\IZ_5'$, which acts by permuting the homogeneous coordinates of $\IP^4$.  It is well
known that there is a family of smooth hypersurfaces invariant under $\IZ_5{\times}\IZ_5'$,
giving rise to the quotient $X^{1,5} = X^{1,101}/\IZ_5{\times}\IZ_5'$.

Now all five fixed points of $\IZ_5$ are identified by the $\IZ_5'$ action, so if we look
for hyperconifold transitions from $X^{1,5}$, we get just one,
\begin{equation*}
    X^{1,5} \rightsquigarrow X^{5,4}~,
\end{equation*}
where the new manifold has fundamental group $\IZ_5$, and is a free quotient of
$X^{21,16}$ from above.  Once again, we have found a brand new manifold, in fact
the first one known with Hodge numbers $\hodgenos = (5,4)$.

It is clear that there are many new manifolds, some with quite small
Hodge numbers, waiting to be found via hyperconifold transitions from known spaces.
No systematic approach to this has been attempted.

\subsection
[Calabi-Yau hypersurfaces in P4(2,1,1,1,1)]
{Calabi-Yau hypersurfaces in $\IP^4_{(2,1,1,1,1)}$} \label{sec:weighted}

It is clear from the previous sections that hyperconifolds do not only occur in the
moduli space of multiply-connected Calabi-Yau manifolds.  More generally, they
can occur in families of varieties which are complete intersections in an ambient
space with orbifold singularities, where a generic member of the family does
not intersect the singularities.  As an example, we will consider Calabi-Yau
hypersurfaces in the weighted projective space $\IP^4_{(2,1,1,1,1)}$.

If the homogeneous coordinates $(z_0, z_1, z_2, z_3, z_4)$ are assigned
weights $(2,1,1,1,1)$, then the resulting weighted projective space has a
$\IZ_2$ orbifold singularity at the point $(1,0,0,0,0)$.  This can be seen by
considering the affine patch $z_0 \neq 0$, and noticing that there are two
choices of rescaling parameter which set $z_0 \to 1$; they are
$\pm\frac{1}{\sqrt{z_0}}$.  The corresponding local coordinates are
therefore subject to the identification
$(y_1, y_2, y_3, y_4) \sim (-y_1, -y_2, -y_3, -y_4)$.

The family of Calabi-Yau hypersurfaces in this space are cut out by
degree six (weighted) homogeneous polynomials.  A generic such polynomial
can, by a $GL(4,\IC)$ transformation on the last four coordinates, be put in
the form
\begin{equation*}
    f = \a_0\, z_0^3 + z_0^2(z_1 z_4 - z_2 z_3) + \ldots~.
\end{equation*}
The corresponding smooth hypersurfaces are simply-connected, by Theorem
1.6 of \cite{Batyrev:2005jc}, and the Hodge numbers are
$\hodgenos = (1,103)$.

In the local coordinates near the orbifold point, $f$ is just
\begin{equation*}
    f = \a_0 + (y_1 y_4 - y_2 y_3) + \ldots~,
\end{equation*}
so that over the distinguished locus in moduli space given by $\a_0 = 0$,
the family develops a $\IZ_2$-hyperconifold singularity.  As mentioned
earlier, this singularity can be resolved by blowing up the orbifold point in
the ambient space, taking us to a new family of smooth simply-connected
Calabi-Yau threefolds, with Hodge numbers $\hodgenos = (2,102)$.  This is
easily confirmed by use of the toric formalism.

This is an example of a hyperconifold transition between two
simply-connected families, which furthermore does not belong to a series
of such transitions starting with a multiply-connected manifold (if so, the
Hodge number $h^{1,1}$ would have to be larger).

It should also be mentioned that hyperconifolds are not the only possibility
in analogous situations.  In some cases, a singularity will arise which is a
quotient of a hypersurface singularity more severe than a node.  The reader
can see an example of this by considering Calabi-Yau hypersurfaces in
$\IP^4_{(4,1,1,1,1)}$.

%%%
\section{Hyperconifolds in Type IIB string theory}\label{sec:string}

Singular Calabi-Yau varieties are particularly interesting in the context of
string compactification where, contrary to intuition, they often give rise to a
consistent physical theory.  It has been known since the pioneering work of
\cite{Dixon:1985jw,Dixon:1986jc} that orbifold singularities can be understood
in the context of string perturbation theory, whereas conifold singularities represent
singularities of the worldsheet theory.  However, in non-perturbative Type IIB string
theory, the conifold singularity is resolved by the effects of light D-brane states
\cite{Strominger:1995cz}.  Furthermore, when it is mathematically possible to carry out
a conifold \emph{transition} to a new Calabi-Yau manifold, this manifests in the physics
as a new branch of the low-energy moduli space \cite{Greene:1995hu}.  It was
suggested in \cite{Davies:2009ub} that a similar story should hold for hyperconifolds,
and we will now show that this is indeed the case.  The following argument is closely
modelled on that of \cite{Greene:1995hu,Strominger:1995cz}, and also uses the insights
of \cite{Gopakumar:1997dv} about D-branes wrapped on multiply-connected cycles.
Much of what follows is well known, but is included in order to give a relatively
self-contained account.

\subsection{The conifold}

For expository reasons, we will consider the case where a hyperconifold singularity
arises in a space $X = \widetilde X/\IZ_N$, so that we can first consider the conifold
singularity which occurs on the covering space $\widetilde X$.

On a Calabi-Yau threefold $\widetilde X$, the homology group $H_3(\widetilde X,\IZ)$
has a symplectic basis (with respect to the intersection form, which is necessarily
symplectic) represented by three-cycles $\{A^I, B_I\}_{I=1,\ldots,h^{2,1}(\widetilde X)+1}$.
The complex structure moduli space of $\widetilde X$ admits complex homogeneous
coordinates $Z^I$, and holomorphic `functions' $F_I$ defined in terms of the
holomorphic three-form $\O$ by \cite{Candelas:1990pi}
\begin{equation*}
    Z^I = \int_{A^I} \O ~,~~ F_I = \int_{B_I} \O ~.
\end{equation*}
The moduli space metric is K\"ahler, with K\"ahler potential
\begin{equation} \label{eq:Kahlerpot}
    K = -\log\left[\ii\left(\overline{Z}^I F_I - Z^I \overline{F}_I\right)\right]~.
\end{equation}
The low-energy dynamics of the complex structure moduli fields is that of a
non-linear sigma model with metric following from this potential.

There are harmonic three-forms $\{\a_I, \b^I\}_{I=1,\ldots,h^{21}(\widetilde X)+1}$
on $\widetilde X$ which are dual to the above cycles, and also related to each other
by the Hodge star operator, $\b^I = \ast \a_I$.
The IIB theory contains a four-form potential $C^{(4)}$, with self-dual five-form field
strength\footnote{In a background where all antisymmetric tensor fields are
set to zero, as we consider here, we have simply $F^{(5)} = dC^{(4)}$.}
$F^{(5)} = \ast F^{(5)}$.  Upon compactification, this gives rise to a number of massless
$U(1)$ gauge fields, one for each of the harmonic three-forms, via a Kaluza-Klein
reduction:
\begin{equation} \label{eq:KK}
    C^{(4)} = \sum_I \left( C^I\wedge\a_I + \widetilde C_I\wedge \b^I \right)
        + \dots~.
\end{equation}
The $C_I$ and $\widetilde C^I$ are massless four-dimensional vector fields, and
the self-duality constraint on $F^{(5)}$ implies the usual four-dimensional electric-magnetic
duality relation $d\widetilde C_I = \ast dC^I$.  These vector fields pair up with the
moduli fields $Z^I$ to give the bosonic content $(C^I, Z^I)$ of $h^{2,1}(\widetilde X)+1$
$\cN=2$ vector multiplets, corresponding to the gauge group $U(1)^{h^{2,1}(\widetilde X)+1}$.

Now suppose we approach a point in complex structure moduli space where $\widetilde X$
develops a conifold singularity.  At the conifold point, a particular three-sphere
vanishes, and we will assume that this is the cycle $A^1$.  We chose our basis of
harmonic three-forms so that only $\a_1$ has a non-zero integral over this cycle,
\begin{equation*}
    \int_{A^1} \a_1 = 1~.
\end{equation*}
A D3-brane couples electrically to $C^{(4)}$, so the action for such a brane which is
wrapped around $A^1$ and follows a worldline $\g$ in the four non-compact
dimensions contains the term
\begin{equation*}
    I_{D3} \supset \int_{A^1\times\g} C^{(4)} = \int_{A^1}\a_1\int_\g C_1 = \int_\g C_1 ~.
\end{equation*}
In the four-dimensional theory, these states therefore manifest as a hypermultiplet
carrying unit electric charge under the $U(1)$ corresponding to the gauge field $C_1$.
The mass of this hypermultiplet saturates a BPS bound coming from the $\cN=2$
supersymmetry algebra \cite{Strominger:1995cz},
\begin{equation*}
    M_{D3} ~\propto~ |Z^1| = \left|\int_{A^1}\O\right| ~\to~ 0~.
\end{equation*}
At the conifold point, then, this hypermultiplet becomes massless, and so should be
included in the low-energy theory.  If instead it is integrated out, it exactly reproduces
the classical singularity of the moduli space, via a divergent one-loop contribution to
$F_1$ \cite{Seiberg:1994rs},
\begin{equation} \label{eq:divergence}
    F_1 \sim \text{const.} + \frac{1}{2\pi\ii} Z^1\log Z^1 ~.
\end{equation}
If this is substituted into \eref{eq:Kahlerpot}, it is easily seen that the moduli space
metric becomes singular at $Z^1 = 0$.  However, this is now seen to be merely an
artifact of integrating out massless states.

The above is a telegraphic account of Strominger's description of conifold singularities
in type IIB string theory.  Now we will ask what happens when the conifold singularity of
$\widetilde X$ lies over a $\IZ_N$-hyperconifold on $X = \widetilde X/\IZ_N$.

\subsection{Hyperconifolds and their resolutions}

First, we observe that the moduli space of $X$ is just a subspace of that of $\widetilde X$,
and inherits its K\"ahler geometry.  Since by assumption the cycle $A^1$ is mapped to
itself by the $\IZ_N$ action, $Z^1$ is a good coordinate on this subspace, and we get
exactly the same singularity implied by \eref{eq:divergence}.  If the hyperconifold
singularity is to make physical sense, we must find states on $X$
which become massless at $Z^1 = 0$ and again reproduce \eref{eq:divergence} if
integrated out.

Such states are easy to identify.  We now have a vanishing cycle $A^1/\IZ_N$, which
again can be wrapped by a D3-brane.  But the worldvolume theory of such a brane
contains a $U(1)$ gauge field, so now that the worldvolume has fundamental group
$\IZ_N$, its vacuum becomes $N$-fold degenerate, corresponding to the $N$ choices
of discrete Wilson line \cite{Gopakumar:1997dv}.  So instead of a single massless
hypermultiplet, the theory on the quotient space $X$ contains $N$ such
hypermultiplets.\footnote{We might also wonder about massless states coming from
winding modes of strings which attain zero length on the hyperconifold.  See
\cite{Gopakumar:1997dv} for a nice explanation of why these need not be
considered separately.}

One might expect that these extra states lead to conflict with \eref{eq:divergence},
since each hypermultiplet will give the same contribution to $F_1$.  But this is a little
too hasty.  Equation \eqref{eq:divergence} comes about from a one-loop calculation,
so the contribution of each hypermultiplet is proportional to the square of its charge, and
we need to check whether this changes when passing from $\widetilde X$ to $X$.
When we perform the Kaluza-Klein expansion of $C^{(4)}$ in \eref{eq:KK}, the
normalisation of the resulting kinetic terms for the $C^I$ depends on the normalisation
of the $\a_I$, which is\footnote{We have normalised the $\a_I$ by the condition
$\int_{A^J} \a_I = \d^J_I$.  Since $\b^I$ is Hodge-dual to $\a_I$ and Poincar\'e dual to
$A^I$, we automatically get
\begin{equation*}
    \int_{\widetilde X} \a_I\wedge\ast \a_I = \int_{\widetilde X} \a_I\wedge \b^I = \int_{A^I} \a_I = 1~.
\end{equation*}
}
\begin{equation*}
    \int_{\widetilde X} \a_I \wedge \ast\a_I = 1~.
\end{equation*}
The same condition should hold on $X$, but now we are integrating over only
$1/N$ times the volume.  The harmonic forms in which we expand $C^{(4)}$ on
$X$ should therefore be $\a'_I = \sqrt{N}\a_I$ (where $I$ now ranges over only those
values for which $\a_I$ is invariant under the group).  As such, the charge of a D3-brane
wrapped on $A^1/\IZ_N$ is
\begin{equation*}
    \int_{A^1/\IZ_N} \!\a'_1 ~=~ \frac{1}{N}\int_{A^1} \!\sqrt{N}\,\a_1 ~=~ \frac{1}{\sqrt{N}}~.
\end{equation*}
There are $N$ such hypermultiplets, so when integrated out they give
\begin{equation*}
    F_1 \sim \text{const.} + N{\times}\left(\!\frac{1}{\sqrt{N}}\right)^2 \frac{1}{2\pi\ii}Z^1\log Z^1 ~,
\end{equation*}
which agrees with \eref{eq:divergence}.  We conclude that hyperconifold
singularities are smoothed by the presence of massless D-brane states, just like the
familiar case of the conifold.

In this paper, and in \cite{Davies:2009ub}, it has been shown that hyperconifolds
can be resolved to pass to a new Calabi-Yau manifold.  Since we now know that
the singularity itself is physically innocuous, we should expect that the theory develops
a new branch of moduli space corresponding to its resolution.  This is true, and the
process is completely analogous to the conifold case, discussed in
\cite{Greene:1995hu}.

First recall that each hypermultiplet contains two complex scalars, each charged under
the $U(1)$ gauge group, so at the hyperconifold point the theory develops $4N$ new
massless scalar degrees of freedom, transforming non-trivially under the $U(1)$.  We
now argue that some of these are flat directions, corresponding to the the new K\"ahler
parameters of the resolution.

The $\cN=2$ vector multiplet of interest contains one real and one complex auxiliary
scalar, which in $\cN=1$ language are respectively the $D$-term associated with the
vector $C^1$, and the $F$-term associated with the complex modulus $Z^1$.  At the
hyperconifold point $Z^1 = 0$, the vacuum conditions become just $D = F = 0$.  These
auxiliary fields are functions of the scalar components of the hypermultiplets charged
under $C^1$, so we get three real conditions on these scalars.  There is also a
one-parameter group of gauge rotations, which removes another degree of freedom.
So we do indeed get a new $4N - 4 = 4(N-1)$-dimensional branch of moduli space,
parametrised by $N-1$ hypermultiplets coming from the new massless states.  Giving
vacuum expectation values to these fields Higgses the $U(1)$ and gives mass to
both $C^1$ and $Z^1$.  In this way it corresponds to moving into the moduli space
of the resolution of the hyperconifold; the new hypermultiplets are identified with
the new K\"ahler parameters, and the fact that $Z^1$ becomes massive corresponds to
the loss of a single complex structure parameter.

\subsection*{Acknowledgements}

I would like to thank Mark Gross for helpful correspondence.
This work was supported by the Engineering and Physical Sciences
Research Council [grant number EP/H02672X/1].

\newpage

\bibliographystyle{utphys}
\bibliography{references}

\end{document}